# Superstatistics – a Quantum Generalization

A. K. Rajagopal, Department of Computer Science, George Mason University, Fairfax, VA 22030 and Inspire Institute Inc., McLean, VA 22101

**ABSTRACT:** A quantum mechanical generalization of superstatistics is presented here based on the positive operator valued measure transformation property of the system density matrix. This procedure reveals that the origin of the fluctuating factors occurring in the derivation of the superstatistics lies in the choice of the transformation operators governing the dynamics of fluctuations. This generalization addresses situations such as nanosystems based on quantum devices (e.g., superconducting devices, single electron transistors, etc,) operating at low temperatures where cognizance of quantum fluctuations is essential.

Superstatistics [1, 2] was conceived as an effective description for describing classical nonequilibrium systems with some varying intensive parameter such as temperature in nanosystems [3]. A review of recent developments and applications to hydrodynamics, astrophysics, etc. may be found in [4]. All of these discussions are based on classical statistical considerations of systems obeying classical mechanical principles. The dynamical underpinning of these discussions is the Langevin description of fluctuations leading to the probability density obeying a corresponding Fokker-Planck equation. Basically, the stationary probability density of the nonequilibrium system for example, arising out of the usual Boltzmann factors, $\exp -\beta E$, is averaged over the fluctuating inverse temperatures, $f(\beta)$:

$$B(E) = \int_0^\infty d\beta \, f(\beta) \exp - \beta E \tag{1}$$

Several forms of superstatistical distributions are phenomenologically proposed depending on the type of fluctuations considered to be pertinent to the physics of the problem. For example, $f(\beta)$ may be the $\chi^2 -$ distribution arising from fluctuation of temperature as in nanosystems [3] lead to power-law type distributions whereas if it is log-normal distribution, it gives rise to a different type [1,2], and so on. In fact, the dynamical considerations of grain growth in [5] leads to a log-normal distribution for the grains from detailed considerations of the Fokker-Planck dynamics of grain growth in two dimensions.

The purpose of the present communication is to address similar situations that call for a quantum statistical mechanical description as with nano-device systems operating at low temperatures, for example. This development requires consideration of the density matrix of the system subject to quantum mechanical fluctuating influences on it by its surroundings. The stationary solution of the Fokker-Planck equation considered in the classical description is here replaced by the stationary solution of the system density matrix under the influence of the fluctuations. We will outline this development in general terms and make connections to real systems afterwards.

Consider the system under consideration described by its density matrix $\hat{r}(A)$. The influence of fluctuating forces on this is represented in the Krauss form [6]:

$$\hat{r}(A) = \sum_a \hat{V}_a \hat{r}_0(A) \hat{V}_a^+, \quad \sum_a \hat{V}_a^+ \hat{V}_a = \hat{I} \qquad (2)$$

Here the set $\{\hat{V}_a\}$ are the nonnegative operators (in general there are more than one such operator and are time-dependent when considering the dynamical evolution) called positive operator valued measure (POVM). (If there is only one such operator, we have the familiar unitary transformation.) They represent the effects of the influence of the surroundings on the system A under consideration and are such that the basic properties of hermiticity, positive semi-definiteness, and traceclass of the density matrix are preserved. $\hat{r}_o(A)$ is the given density matrix of system A initially. For simplicity of presentation, we have used a discrete state representation; the continuous state version in the above expressions would involve integration in place of summation over the states. $\sum_a \to \int d\mathbf{m}_a$.

This representation is quite general from which approximate equation such as the Lindblad evolution, governing the density matrix is derived [6]. The stationary density matrix of eq.(1) is the object of interest corresponding to the stationary solution of the Fokker – Planck equation mentioned above and here we will not denote it by a different symbol for simplicity of presentation. Since the density matrices are hermitian operators, they admit of a diagonal spectral representation of the form:

$$\hat{r}(A) = \sum_a |a\rangle P(a)\langle a|, \quad \hat{r}_0(A) = \sum_{a_0} |a_0\rangle p_0(a_0)\langle a_0| . \qquad (3)$$

Here $P(a)$ corresponds to the stationary probability of the system and $p_0(a_0)$ to the Boltzmann – like factor for the initial state of the system. By inserting these expressions in eq.(3), we have finally the relationship of the form

$$P(a) = \sum_{a_0} \Pi(a|a_0) p_0(a_0), \quad \Pi(a|a_0) = \sum_a |\langle a|\hat{V}_a|a_0\rangle|^2 \qquad (4)$$

We observe that

$$\sum_a \Pi(a|a_0) = \sum_a \langle a_0|\hat{V}_a^+ V_a|a_0\rangle = 1 \qquad (5)$$

after using the condition in eq.(2) on the POVM operators. Note that the sum over the other indices, $\{a_0\}$, will not lead to one as in eq.(5) unless the POVM operators are unital, $\sum_a \hat{V}_a \hat{V}_a^+ = I$.

The result in eq.(4) corresponds to the "Quantum Superstatistics" expressing the stationary distribution as a sum over the initial distribution weighted over by the suitable factor explicitly shows how the environment is influencing the final distribution.

It may be noted in passing that the von Neumann entropy of the stationary state may be either increase or decrease from that of the initial distribution. This is an important property of the POVMs.

The applicability of such a quantum description is relevant for describing nanosystems at sufficiently low temperatures where quantum effects are dominant as in devices based on superconductivity, semiconductor quantum dots, single electron transistors, and the like [7].

**REFERENCES:**
[1] C. Beck, Phys. Rev. Lett. **87**, 180601 (2001)
[2] C. Beck and E. G. D. Cohen, Physica A**322**, 267 (2003)
[3] A. K. Rajagopal, C. S. Pande, and S. Abe, cond-mat/0403738; to appear in the proceedings of the Indo-US Workshop on *"Nanoscale materials: From Science to Technology"*, (Nova Publications, New York ,2006)